%% file: main_V2.tex
\documentclass[lettersize,journal]{IEEEtran}

\usepackage[T1]{fontenc}
\usepackage{amsmath,amssymb,amsfonts}
\usepackage{bm}
\usepackage{graphicx}
\usepackage{cite}
\usepackage{booktabs}
\usepackage{multirow}
\usepackage{array}
\usepackage{url}
\usepackage{textcomp}
\usepackage{stfloats}
\usepackage{verbatim}
\usepackage{xcolor}
\usepackage{balance}
\usepackage{placeins}
\usepackage{siunitx}

\usepackage[caption=false,font=normalsize,labelfont=sf,textfont=sf]{subfig}

\usepackage{tikz}
\usetikzlibrary{positioning, fit, backgrounds, arrows.meta, shadows, calc}

\usepackage{algorithm}
\usepackage{algpseudocode}

\usepackage[hidelinks,breaklinks]{hyperref}

\graphicspath{{journal_figures/}{author_photos/}}

\newif\ifdraftcomments
\draftcommentstrue

\ifdraftcomments
\usepackage[textsize=tiny]{todonotes}
\usepackage[normalem]{ulem}

\newcommand{\rd}[1]{\textcolor{red}{\sout{#1}}}

\newcommand{\authcomment}[1]{\par\noindent\textcolor{red}{\emph{Author comment: #1}}\par}
\newcommand{\authornote}[1]{\par\noindent\textcolor{blue}{\emph{Author to fill: #1}}\par}
\else
\newcommand{\todo}[2][]{}
\newcommand{\rd}[1]{}

\newcommand{\authcomment}[1]{}
\newcommand{\authornote}[1]{}
\fi

\definecolor{figBlueD}{HTML}{356F9E}
\definecolor{figBlueL}{HTML}{F2F7FC}

\definecolor{figTealD}{HTML}{3E8D88}
\definecolor{figTealL}{HTML}{F1FAF8}

\definecolor{figAmberD}{HTML}{C4873C}
\definecolor{figAmberL}{HTML}{FFF7EA}

\definecolor{figPlumD}{HTML}{8A528B}
\definecolor{figPlumL}{HTML}{FAF3FB}

\definecolor{figGreenD}{HTML}{5C9563}
\definecolor{figGreenL}{HTML}{F2FAF2}

\definecolor{figGoldD}{HTML}{A68B32}
\definecolor{figGoldL}{HTML}{FFF9E8}

\definecolor{figSlateD}{HTML}{4F5B66}
\definecolor{figSlateL}{HTML}{F5F7F8}

\definecolor{figInk}{HTML}{262626}

\newcommand{\R}{\mathbb{R}}
\newcommand{\y}{\mathbf{y}}
\newcommand{\uvec}{\mathbf{u}}
\newcommand{\x}{\mathbf{x}}
\newcommand{\hvec}{\mathbf{h}}
\newcommand{\phivec}{\boldsymbol{\phi}}
\newcommand{\psivec}{\boldsymbol{\psi}}
\newcommand{\Ical}{\mathcal{I}_{\mathrm{cal}}}
\newcommand{\diag}{\mathrm{diag}}

\makeatletter
\g@addto@macro\normalsize{%
	\setlength{\abovedisplayskip}{4pt plus 1pt minus 1pt}%
	\setlength{\belowdisplayskip}{4pt plus 1pt minus 1pt}%
	\setlength{\abovedisplayshortskip}{0pt plus 1pt}%
	\setlength{\belowdisplayshortskip}{3pt plus 1pt minus 1pt}%
}
\makeatother
\algrenewcommand\algorithmicrequire{\textbf{Input:}}
\algrenewcommand\algorithmicensure{\textbf{Output:}}

\usepackage{etoolbox}
\AtBeginEnvironment{algorithm}{%
	\vspace{5pt}
	\setlength{\abovecaptionskip}{0pt}%
	\setlength{\belowcaptionskip}{0pt}%
}

\AtBeginEnvironment{algorithmic}{%
	\footnotesize
	\setlength{\topsep}{0pt}%
	\setlength{\partopsep}{0pt}%
	\setlength{\itemsep}{0pt}%
	\setlength{\parsep}{0pt}%
}

\begin{document}
	
	\title{Data-Driven Generator Transient Prediction for Digital Twin Decision Support\\ \vspace{5mm}
	\footnotesize {The views expressed are those of the author and do not reflect the official policy or position \\of the Department of War or the U.S. Government.}
}
	
	\author{
		Emad~Sadeghi,~\IEEEmembership{Student Member,~IEEE},
		Emerson~Miller,~\IEEEmembership{Student Member,~IEEE},\\
		Kerry~Sado,~\IEEEmembership{Senior Member,~IEEE},
		and Adel~Nasiri,~\IEEEmembership{Fellow,~IEEE}
		\thanks{The authors are with the Department of Electrical Engineering, University of South Carolina, Columbia, SC, USA. Corresponding author: Adel Nasiri (e-mail: nasiri@sc.edu).}
		\thanks{This work was supported by the Office of Naval Research under contracts No.\,N00014-23-C-1012 and No.\,N00014-24-C-1301. DISTRIBUTION STATEMENT A. Approved for public release: distribution is unlimited. Approved, DCN\# 2026-6-30-2452.}
	}
	
	\markboth{\fontsize{4.5}{5}\selectfont This work has been submitted to the IEEE for possible publication. Copyright may be transferred without notice, after which this version may no longer be accessible.}%
	{Sadeghi \MakeLowercase{\textit{et al.}}: Data-Driven Generator Transient Prediction for Digital Twin Decision Support}
	
	\maketitle
	
	\pagenumbering{gobble}	
	
	\bstctlcite{BSTcontrol}

\begin{abstract}This paper develops a calibrated transient-forecasting surrogate model for generator digital twin (DT) decision support that evaluates planned active- and reactive-power load commands before they are applied. The proposed event-conditioned Hankel Dynamic Mode Decomposition with Control (Hankel-DMDc) model combines delay-coordinate lifting, command-event memory features, and an event-weighted Hankel basis so that sparse load-transition dynamics influence the reduced representation and fitted dynamics. This design targets intervals where voltage/frequency deviations and recovery behavior determine whether a candidate load command keeps the system within acceptable limits. To provide operator-facing confidence information, a split-conformal calibration layer is applied to the frozen surrogate model to form event-conditioned joint prediction bands for voltage and frequency. The experimental results show event-window root-mean-square errors of 1.058~V and 0.155~Hz. For a nominal 90\% target, the bands attain 90.17\% pointwise joint voltage-frequency coverage, with mean band widths of 3.55~V and 0.566~Hz. A 50-s open-loop rollout is computed in 181~ms on a single CPU core, approximately 280 times faster than real time, with forecast accuracy evaluated over horizons up to 5~s. These results demonstrate a computationally efficient advisory framework for generator DTs that combines transient prediction with calibrated uncertainty.
\end{abstract}

\begin{IEEEkeywords}
	Digital twins, generator transient forecasting, reduced-order modeling, uncertainty quantification.
\end{IEEEkeywords}
	
	\input{sections/01_introduction}
	\input{sections/02_background}
	\input{sections/03_forecaster}
	\input{sections/04_hardware_protocol}
	\input{sections/05_results_deterministic}

	\input{sections/06_conformal_reliability}
	\input{sections/09_discussion}
	\input{sections/11_conclusion}
	\input{sections/12_acknowledgment}

\FloatBarrier
\bibliographystyle{IEEEtran}
\bibliography{references}

\vspace{0.75\baselineskip}

\newcommand{\compactbio}[3]{%
	\par\noindent
	\begin{minipage}[t]{\columnwidth}
		\begin{minipage}[t]{0.82in}
			\vspace{0pt}
			\includegraphics[width=0.75in,height=0.95in,clip,keepaspectratio]{#1}
		\end{minipage}
		\hfill
		\begin{minipage}[t]{\dimexpr\columnwidth-0.92in\relax}
			\vspace{0pt}
			\footnotesize
			\textbf{#2} #3
		\end{minipage}
	\end{minipage}
	\par\vspace{0.65\baselineskip}
}

\end{document}

%% file: sections/01_introduction.tex
\section{Introduction}
\IEEEPARstart{G}{enerator} digital twins (DTs) can support dynamic power system operation by forecasting whether planned load commands will violate voltage and frequency limits before those commands are applied. This capability is especially important in islanded or weak-grid systems where load changes can lead to more significant transients\cite{Data_Center, StabilityChallenges}. If these transients exceed operating limits, protection systems or controls may disconnect equipment, interrupting critical loads.

The practical value of a generator DT depends on balancing model fidelity with computational efficiency. Detailed physics-based models, including differential-algebraic-equation and port-Hamiltonian formulations, provide high-fidelity descriptions of generator dynamics and remain essential for planning, controller design, and protection studies \cite{kundur1994}. However, their operational use in a DT can be limited by incomplete parameter knowledge, unmeasured internal states, proprietary controller logic, and changes caused by aging or retuning~\cite{DAE, PortHamiltonian}. Even when a detailed high-fidelity physics-based model is available, repeated look-ahead evaluation can be computationally expensive if small integration time steps or detailed control models are required. These limitations motivate a complementary data-driven approach that learns predictive system dynamics from operating data while remaining computationally efficient enough for DT forecasting \cite{Data_Driven}.

Motivated by these limitations, this paper develops a data-driven surrogate model for generator transient prediction that is conditioned on the commanded load trajectory \cite{sadeghi_ecce2025}. Using this command information together with a recent history of measured voltage and frequency, the model forecasts the terminal response over a prescribed look-ahead horizon. Its computationally efficient surrogate structure enables extended scenario evaluation while preserving the accuracy needed to resolve the voltage and frequency transients that determine operational feasibility. A key challenge, however, is that the most operationally important intervals are often sparsely represented in the measured data. Load steps and pulses occupy only a small fraction of the dataset, yet they determine voltage/frequency deviations, recovery rates, and margins for keeping converters online \cite{ogiesobaeguakun2026highfidelity}. A model identified from unweighted data can therefore be dominated by steady-state behavior. To address this issue, the proposed method makes both the delay-coordinate basis and the model-selection objective event-sensitive, placing emphasis on command transitions, output transients, and recovery shape.

For an operator-facing DT, however, accurate nominal forecasts are not enough on their own. A point prediction may track nominal behavior well while still being insufficient for decision support if it does not provide a calibrated measure of forecast error \cite{DT_Credibility}. Such uncertainty information helps determine when a forecast should be trusted and allows operators to interpret the risk associated with candidate load scenarios. The proposed framework therefore augments the deterministic event-conditioned Hankel Dynamic Mode Decomposition with Control (Hankel-DMDc) surrogate model with conformal uncertainty bands. These bands are calibrated jointly over voltage and frequency because many operating decisions require simultaneous coverage of both quantities rather than separate marginal coverage for each output \cite{fan2025interpretable}.

Taken together, these developments extend the deterministic surrogate model into a transient-focused forecasting framework with calibrated uncertainty for DT decision support. The main contributions of this work are: (i) an event-aware command-memory representation for improved transient prediction via explicit encoding of command-triggered dynamics; (ii) an event-weighted Hankel basis that prioritizes sparse, high-impact command transitions and enhances sensitivity to transient conditions; (iii) a transient-sensitive model-selection criterion balancing average, event-specific, peak, and derivative errors to better reflect operational performance; and (iv) hierarchical event-conditioned joint conformal prediction bands that provide calibrated, simultaneous uncertainty estimates for voltage and frequency to support operator-facing decision-making. Building on \cite{sadeghi_ecce2025}, which established Hankel-DMDc as a deterministic reduced-order generator surrogate model under pulse loads, this work advances the approach by adding event-aware identification, transient-focused model selection, and calibrated joint uncertainty quantification for more reliable, actionable DT prediction.

%% file: sections/02_background.tex
\vspace{-2mm}
\section{Background and Related Work}
Although substantial work exists in DTs and generator modeling, there remains a clear tradeoff between detail, data dependence, and computational cost in models used to predict generator transients. Existing approaches navigate this tradeoff from different perspectives, but practical, data-driven solutions for fast and reliable look-ahead prediction remain limited.
\vspace{-2mm}
\subsection{Digital Twins and Generator Surrogate Modeling}
DTs for power and energy systems combine measurement streams with computational models and operational context to support monitoring, prediction, fault analysis, and decision support \cite{bazmohammadi2022}. Prior work has emphasized DT applications in microgrids and energy systems with local generation, power-electronic interfaces, storage, and flexible loads, where online models can support operating decisions under changing conditions \cite{bazmohammadi2022}. For the present work, the DT function is short-horizon look-ahead prediction of generator terminal voltage and frequency under candidate load-command trajectories.

The literature on generator transient prediction generally spans three broad modeling paradigms: physics-based model reduction, grey-box identification, and data-driven surrogate modeling. Physics-based reduction methods, including balanced truncation, Loewner interpolation, and structure-preserving port-Hamiltonian formulations, reduce model complexity while retaining stability and energy-structure properties \cite{antoulas2005,beattie2009}. These approaches, however, typically require an available differential-algebraic-equation model and detailed generator parameters, which can increase computational burden. This requirement can limit their use when parameters are uncertain, controller details are proprietary, internal states are unmeasured, or repeated look-ahead evaluations are needed.

Grey-box and measurement-based identification methods reduce the dependence on complete machine models by estimating selected parameters, states, or input-output relationships from disturbance data. Examples include state-estimation-based parameter identification, autoregressive-with-exogenous-input (ARX) models, Hankel-matrix fitting, online learning, and symbolic-regression-based refinement to recover or improve generator dynamic models~\cite{aghamolki2015,alassaf2022,Symbolic_Regression}. More flexible learning-based surrogates, including recurrent networks, operator-learning models, neural differential-equation models, and graph-based predictors, can represent complex dynamics with fewer prescribed model equations. Their use in operator-facing DTs, however, requires representative large training datasets, validation, and uncertainty assessment \cite{nandanoori2022,moya2023}.

These modeling approaches highlight the tradeoff between physical detail, data dependence, and computational cost in generator transient prediction. Motivated by this tradeoff, the work here develops a hardware-validated surrogate model for DT look-ahead forecasting, improving the balance between computational efficiency and transient-event accuracy.
\vspace{-2mm}
\subsection{Koopman-Based Dynamic Surrogates}
\begin{figure*}[t]
	\centering
	\scalebox{0.82}{%
		\begin{tikzpicture}[
			font=\rmfamily\footnotesize,
			>=Latex,
			box/.style={
				draw,
				line width=0.9pt,
				rounded corners=4pt,
				minimum width=4.00cm,
				text width=3.75cm,
				minimum height=0.88cm,
				align=center,
				text centered,
				text=figInk,
				inner xsep=5pt,
				inner ysep=3pt,
				drop shadow={
					shadow xshift=0.8pt,
					shadow yshift=-0.8pt,
					opacity=0.12
				}
			},
			edgebox/.style={
				box,
				minimum width=3.10cm,
				text width=2.85cm
			},
			data/.style={box, fill=figBlueL, draw=figBlueD},
			dataedge/.style={edgebox, fill=figBlueL, draw=figBlueD},
			prep/.style={box, fill=figTealL, draw=figTealD},
			model/.style={box, fill=figAmberL, draw=figAmberD},
			trust/.style={box, fill=figPlumL, draw=figPlumD},
			trustedge/.style={edgebox, fill=figPlumL, draw=figPlumD},
			scenario/.style={edgebox, fill=figGreenL, draw=figGreenD},
			display/.style={edgebox, fill=figGoldL, draw=figGoldD},
			arr/.style={->, line width=0.9pt, draw=figInk!95},
			dash arr/.style={
				->,
				line width=0.8pt,
				dashed,
				draw=figInk!70
			},
			transfer label/.style={
				fill=figSlateL!0,
				fill opacity=0.95,
				text opacity=4,
				text=figInk!95,
				font=\rmfamily\scriptsize\itshape,
				inner xsep=2.5pt,
				inner ysep=0.7pt
			},
			container/.style={
				draw=figSlateD!95,
				dashed,
				rounded corners=6pt,
				inner sep=9pt,
				fill=figSlateL!85
			}
			]
			
			\matrix[
			row sep=1.55cm,
			column sep=0.55cm,
			ampersand replacement=\&
			] {
				\node[dataedge] (a)
				{Measured Generator\\ Terminal Data};
				\&
				\node[prep] (b)
				{Signal Preprocessing and\\ Command-Event Features};
				\&
				\node[model] (c)
				{Event-Conditioned\\ Hankel-DMDc Surrogate};
				\&
				\node[trustedge] (d)
				{Joint Voltage/Frequency\\ Band Calibration};
				\\
				\node[scenario] (e)
				{Planned $P/Q$ Scenario:\\ Step and Pulse};
				\&
				\node[model] (f)
				{Open-Loop Surrogate Rollout\\ Under Planned Setpoints};
				\&
				\node[trust] (g)
				{Voltage/Frequency Forecast\\ With Joint Conformal Band};
				\&
				\node[display] (h)
				{Operator-Facing\\ Digital-Twin Display};
				\\
			};
			
			\draw[arr] (a) -- (b);
			\draw[arr] (b) -- (c);
			\draw[arr] (c) -- (d);
			
			\draw[arr] (e) -- (f);
			\draw[arr] (f) -- (g);
			\draw[arr] (g) -- (h);
			
			\draw[dash arr]
			(c.south) .. controls +(-0.25,-0.78) and +(0.15,0.60) ..
			node[
			pos=0.55,
			above=2.2pt,
			rotate=8,
			transfer label
			] {Frozen Surrogate}
			(f.north);
			
			\draw[dash arr]
			(d.south) .. controls +(0.25,-0.95) and +(0.20,0.75) ..
			node[
			pos=0.55,
			above=3pt,
			rotate=8,
			transfer label
			] {Calibrated Band Table}
			(g.north);
			
			\begin{scope}[on background layer]
				\node[
				container,
				fit=(a)(b)(c)(d),
				label={
					[font=\rmfamily\small\bfseries, text=figInk, yshift=1mm]
					above:Offline Data-Driven Refresh
				}
				] {};
				
				\node[
				container,
				fit=(e)(f)(g)(h),
				label={
					[font=\rmfamily\small\bfseries, text=figInk, yshift=-1mm]
					below:Online Predictive Digital Twin
				}
				] {};
			\end{scope}
			
		\end{tikzpicture}%
	}\vspace{-2mm}
	\caption{Calibrated transient-prediction workflow for the generator digital twin. The surrogate model is updated offline from hardware data and queried online under planned active- and reactive-power load scenarios, returning voltage and frequency forecasts with a joint conformal band.}
	\label{fig:workflow}
\end{figure*}
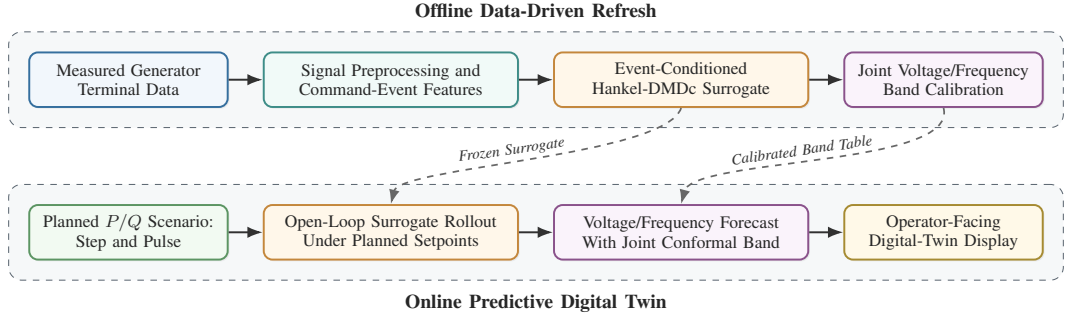

Dynamic mode decomposition (DMD) and its variants provide finite-dimensional linear representations of measured nonlinear dynamics \cite{Colbrook_2023}. Classical DMD is closely connected to Koopman spectral analysis, while DMD with control (DMDc) incorporates exogenous inputs and is therefore well suited for commanded-load scenarios \cite{rowley2009,schmid2010,proctor2016}. Related methods, including extended DMD and controlled Koopman predictors, expand the observable space in which linear predictors are identified \cite{williams2015,korda2018}. When the internal state is not directly measured, Hankel delay-coordinate formulations are particularly useful, as histories of output samples can encode information about hidden dynamics \cite{brunton2017}. These methods have also been applied to power-system stability and transient analysis, including Koopman-mode indicators of swing instability and Koopman-based transient-stability prediction \cite{susuki2012,jafarzadeh2021,choi2024}. The proposed approach leverages the Koopman perspective to identify a delay-coordinate linear surrogate for forecasting voltage and frequency under planned load scenarios, as embedded in the calibrated DT workflow of Fig.~\ref{fig:workflow}.

%% file: sections/03_forecaster.tex
\section{Event-Conditioned Surrogate Model}
\label{sec:method}
The proposed surrogate model is developed in two stages. First, recent terminal measurements are combined with the planned load-command trajectory to generate open-loop voltage and frequency forecasts. Second, the trained surrogate model is held fixed while joint conformal bands are calibrated around the two output trajectories. This section describes the event-conditioned forecasting stage.
\vspace{-2mm}
\subsection{Command-Conditioned Forecasting Map}
Let time be discretized as $t_k = k\Delta t$. At each time step, the measured output is
\begin{equation}
	\y_k =
	\begin{bmatrix}
		y_{V,k} & y_{f,k}
	\end{bmatrix}^{\top}
	\in \R^2,
	\label{eq:y_def}
\end{equation}
where $y_{V,k}$ and $y_{f,k}$ denote the normalized terminal RMS voltage ($V$) and frequency ($f$), respectively. The command applied over the same time step is
\begin{equation}
	\uvec_k =
	\begin{bmatrix}
		u_{P,k} & u_{Q,k}
	\end{bmatrix}^{\top}
	\in \R^2,
	\label{eq:u_def}
\end{equation}
where $u_{P,k}$ and $u_{Q,k}$ are the scaled active ($P$) and reactive power ($Q$) setpoints.

At the forecast starting point $k$, the model is given the recent measurement history and the command sequence covering both the recent history and the prediction horizon. The forecasting map is therefore
\begin{equation}
	\left\{\y_{k-q+1:k}, \uvec_{k-q+1:k+H}\right\}
	\mapsto
	\left\{\widehat{\y}_{k+\ell|k}\right\}_{\ell=1}^{H},
	\label{eq:forecast_map}
\end{equation}
where $q$ is the delay length and $H$ is the forecast horizon. In this formulation, the future portion of the command sequence defines the operator-selected scenario, and the model returns the corresponding open-loop voltage and frequency response.

The forecasts are evaluated in two ways. First, the model is rolled across the full held-out interval to assess overall tracking performance. Since this record is dominated by long steady-state stretches, the full-record view mainly provides a broad consistency check. Second, per-event look-ahead forecasts are generated by restarting the rollout at each detected event and propagating the model over a fixed horizon. This evaluation isolates the transient windows that the model is designed to predict. A horizon of $H_{\mathrm{s}}$ seconds corresponds to $H=\lfloor H_{\mathrm{s}}/\Delta t\rfloor$ samples at the model sample period $\Delta t$.

\subsection{Delay-Coordinate Observable}
A single voltage/frequency sample provides limited information about the underlying dynamic state of the machine. Delay coordinates address this limitation by using recent measurement histories to construct a richer representation of the latent dynamics. For a sequence $\{\y_k,\uvec_k\}_{k=1}^{N}$ and delay length $q$, set $M=N-q+1$ and
\begin{equation}
    \hvec_i=\begin{bmatrix}\y_i^{\top}&\y_{i+1}^{\top}&\cdots&\y_{i+q-1}^{\top}\end{bmatrix}^{\top}\in\R^{2q}.
    \label{eq:hankel_vector}
\end{equation}
Stacking these windows gives the Hankel data matrix
\begin{equation}
	H_y(q)=
	\begin{bmatrix}
		\hvec_1 & \hvec_2 & \cdots & \hvec_M
	\end{bmatrix}^{\top}
	\in\R^{M\times 2q}.
	\label{eq:hankel_matrix}
\end{equation}
Each row ends at the terminal output $\y_{i+q-1}$. In online prediction, this same construction defines the measured-delay block $\hvec_k^{\leftarrow}=[\y_{k-q+1}^{\top},\ldots,\y_k^{\top}]^{\top}$ at time $k$, corresponding to row $i=k-q+1$. This distinction keeps the measured output window used to construct the state separate from the future command sequence that drives the rollout. The Hankel window therefore provides the measured memory through which the reduced model represents internal dynamics that are not directly observed.

\subsection{Event-Aware Command Features}
The command profile contains information beyond the instantaneous power levels. In particular, the transient response depends on recent command changes, including their sign, magnitude, and timing. To encode this information, define
\begin{align}
    \Delta u_{P,k} &= u_{P,k}-u_{P,k-1},\label{eq:duP}\\
    \Delta u_{Q,k} &= u_{Q,k}-u_{Q,k-1},\label{eq:duQ}\\
    S_k &= \sqrt{u_{P,k}^{2}+u_{Q,k}^{2}},\label{eq:S}\\
    \Delta S_k &= S_k-S_{k-1}.\label{eq:dS}
\end{align}
The overall size of a command change is then the scalar
\begin{equation}
    m_k=\left[(\Delta u_{P,k})^2+(\Delta u_{Q,k})^2\right]^{1/2}.
    \label{eq:magnitude}
\end{equation}
To separate the initial transient response from the slower recovery dynamics, the model passes $m_k$ through a bank of exponentially decaying memory channels,
\begin{equation}
	z_{\tau,k}=a_{\tau}z_{\tau,k-1}+(1-a_{\tau})m_k,
	\qquad
	a_{\tau}=\exp\left(-\frac{\Delta t}{\tau}\right),
	\label{eq:event_memory}
\end{equation}
where $z_{\tau,k}$ is the event-memory state associated with time constant $\tau$, and $a_{\tau}$ is the corresponding decay factor. One channel is used for each time constant in the set $\{\tau_1,\ldots,\tau_m\}$, spanning fast to slow recovery behavior. The resulting raw feature vector is
\begin{equation}
\begin{split}
    \phivec_k^{\mathrm{raw}}=[&u_{P,k},u_{Q,k},\Delta u_{P,k},\Delta u_{Q,k},\\
    &|\Delta u_{P,k}|,|\Delta u_{Q,k}|,S_k,\Delta S_k,\\
    &z_{\tau_1,k},\ldots,z_{\tau_m,k}]^{\top}.
\end{split}
\label{eq:phi_raw}
\end{equation}
Every component is standardized with training-set statistics. The dynamics then read these features over a time-lagged window. Let
\begin{equation}
    \mathcal E_q(s)=\{0,s,2s,\ldots,\lfloor(q-1)/s\rfloor s\}\cup\{q-1\}
    \label{eq:lag_set}
\end{equation}
be the lag set with stride $s$, written in increasing order as $\mathcal E_q(s)={e_1,\ldots,e_L}$. Stacking the standardized features over these $L$ lags gives the command-feature vector for row $i$,
\begin{equation}
    \psivec_i=\begin{bmatrix}
    \phivec_{i+e_1}^{\top}&\phivec_{i+e_2}^{\top}&\cdots&\phivec_{i+e_L}^{\top}
    \end{bmatrix}^{\top}.
    \label{eq:psi}
\end{equation}
This construction aligns the command-feature window with the corresponding output Hankel row. The vector $\psivec_i$ therefore provides the reduced model with a compact memory of recent command activity, without requiring future measurements.

\begin{figure*}[t]
	\centering \vspace{-2mm}
	\scalebox{0.82}{%
		\begin{tikzpicture}[
			font=\rmfamily\footnotesize,
			>=Latex,
			box/.style={
				draw,
				line width=0.9pt,
				rounded corners=4pt,
				minimum width=3.85cm,
				text width=3.55cm,
				minimum height=0.92cm,
				align=center,
				text centered,
				text=figInk,
				inner xsep=5pt,
				inner ysep=3pt,
				drop shadow={
					shadow xshift=0.8pt,
					shadow yshift=-0.8pt,
					opacity=0.12
				}
			},
			tallbox/.style={
				box,
				minimum height=1.12cm,
				inner ysep=5pt
			},
			compacttallbox/.style={
				tallbox,
				minimum width=2.8cm,
				text width=2.8cm
			},
			edgebox/.style={
				box,
				minimum width=3.35cm,
				text width=3.10cm
			},
			talledgebox/.style={
				edgebox,
				minimum height=1.12cm,
				inner ysep=5pt
			},
			compacttalledgebox/.style={
				talledgebox,
				minimum width=2.2cm,
				text width=2.4cm
			},
			bluetall/.style={tallbox, fill=figBlueL, draw=figBlueD},
			bluemidtall/.style={compacttallbox, fill=figBlueL, draw=figBlueD},
			bluetalledge/.style={talledgebox, fill=figBlueL, draw=figBlueD},
			orangetall/.style={tallbox, fill=figAmberL, draw=figAmberD},
			orangemidtall/.style={compacttallbox, fill=figAmberL, draw=figAmberD},
			orangetalledge/.style={talledgebox, fill=figAmberL, draw=figAmberD},
			greentall/.style={tallbox, fill=figGreenL, draw=figGreenD},
			greenmidtall/.style={compacttallbox, fill=figGreenL, draw=figGreenD},
			greentalledge/.style={talledgebox, fill=figGreenL, draw=figGreenD},
			graytall/.style={tallbox, fill=figPlumL, draw=figPlumD},
			graytalledge/.style={talledgebox, fill=figPlumL, draw=figPlumD},
			graycompactedge/.style={compacttalledgebox, fill=figPlumL, draw=figPlumD},
			arr/.style={
				->,
				line width=0.9pt,
				draw=figInk!75
			},
			statearr/.style={
				->,
				line width=0.9pt,
				draw=figInk!70
			},
			container/.style={
				draw=figSlateD!95,
				dashed,
				rounded corners=6pt,
				inner xsep=10pt,
				inner ysep=10pt,
				fill=figSlateL!85
			}
			]
			
			\def\flowgap{0.45cm}
			\def\rowsep{1.1cm}
			
			\node[bluetalledge] (a)
			{\textbf{Measured History}\\[2pt]  Voltage/Frequency\\ Window $y_{k-q+1:k}$};
			
			\node[bluemidtall, right=\flowgap of a] (b)
			{\textbf{Hankel Observable}\\[2pt]  Delay Coordinates\\ $h_k\in\mathbb{R}^{2q}$};
			
			\node[orangemidtall, right=\flowgap of b] (c)
			{\textbf{Event-Weighted Basis}\\[2pt]  Transient Command-Change Windows};
			
			\node[orangetalledge, right=\flowgap of c] (d)
			{\textbf{Reduced Coordinate}\\[2pt]  Encoder $x_k=E h_k$ };
			
			\node[graytall, below=\rowsep of d] (g)
			{\textbf{Controlled Update}\\[2pt]  Reduced Dynamics\\ $x_{k+1}=Ax_k+B\psi_k$};
			
			\node[greenmidtall, left=\flowgap of g] (f)
			{\textbf{Event Memory}\\[2pt]  Multi-Scale Command\\ Filters $z_{\tau,k}$};
			
			\node[greentalledge, left=\flowgap of f] (e)
			{\textbf{Planned Command}\\[2pt]  $P,Q,\Delta P,\Delta Q$\\ Scenario};
			
			\node[graycompactedge, right=\flowgap of g] (h)
			{\textbf{Forecast Output}\\[2pt]  Terminal Forecast\\ $\widehat{y}_k=Cx_k$};
			
			\draw[arr] (a) -- (b);
			\draw[arr] (b) -- (c);
			\draw[arr] (c) -- (d);
			
			\draw[arr] (e) -- (f);
			\draw[arr] (f) -- (g);
			\draw[arr] (g) -- (h);
			
			\draw[statearr] (d.south) -- (g.north);
			
			\begin{scope}[on background layer]
				\node[
				container,
				fit=(a)(b)(c)(d),
				label={
					[font=\rmfamily\small\bfseries, text=figInk, yshift=1mm]
					above:Measured Observable Lifting
				}
				] {};
				
				\node[
				container,
				fit=(e)(f)(g)(h),
				label={
					[font=\rmfamily\small\bfseries, text=figInk, yshift=-1mm]
					below:Command-Conditioned Reduced Prediction
				}
				] {};
			\end{scope}
			
		\end{tikzpicture}%
	} 
	\caption{Workflow for the deterministic surrogate model. The top row shows the Hankel lifting step, in which the measured voltage/frequency history is embedded as a Hankel observable and projected onto an event-weighted reduced coordinate. The bottom row shows the command-conditioned rollout, where planned load commands and event-memory features drive the reduced dynamics to produce the terminal voltage/frequency forecast.}
	\label{fig:hankel_lifting}
\end{figure*}
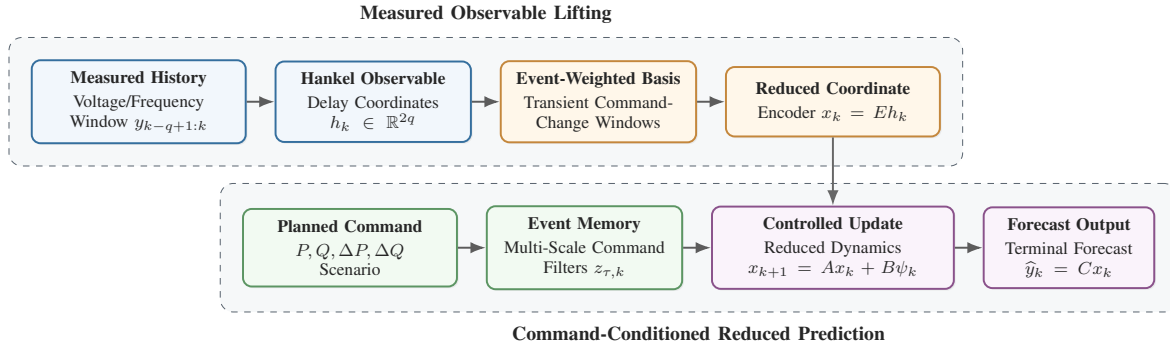

Fig.~\ref{fig:hankel_lifting} summarizes these components: the Hankel observable provides the measured memory, the weighted basis defines the reduced coordinate, and the controlled update advances that coordinate under a planned command.

\subsection{Event-Weighted Hankel Basis}
A standard construction of the Hankel basis uses an unweighted singular value decomposition (SVD) of $H_y(q)$. This basis is optimal for average reconstruction of the training Hankel matrix, but average reconstruction error does not necessarily reflect transient fidelity. To emphasize windows near command transitions and voltage/frequency deviations, the method instead uses a weighted matrix,
\begin{equation}
    \widetilde H_y=W_b^{1/2}H_y=U\Sigma V^{\top}, \, \, \, W_b=\diag(w_1,\ldots,w_M).
    \label{eq:weighted_svd}
\end{equation}
The weighting modifies only the offline subspace identification problem by emphasizing selected Hankel rows during basis construction. During online prediction, the surrogate model is conditioned only on the measured delay window and the planned command profile.

The row weight is defined from three event-sensitive scalar factors. Let $\eta_i^u\in[0,1]$ denote a normalized command-change factor evaluated at the terminal sample of row $i$, let $\eta_i^y\in[0,1]$ denote a training-only transient factor computed from a robust voltage/frequency derivative magnitude, and let $\eta_i^s\in[0,1]$ denote a voltage-sag indicator. These terms are combined to form the basis weight as
\begin{equation}
    w_i=\min\{(1+\alpha_u\eta_i^u)(1+\alpha_y\eta_i^y)(1+\alpha_s\eta_i^s),w_{\max}\}.
    \label{eq:weights_general}
\end{equation}
Because \eqref{eq:weights_general} multiplies the three factors, rows associated with command changes, output transients, and voltage sags receive the largest emphasis when these effects occur together. The cap $w_{\max}$ limits this amplification.
\vspace{-2mm}
\subsection{Controlled Reduced-Order Evolution}
Let $r$ denote the retained rank of the weighted SVD in \eqref{eq:weighted_svd}. Truncating the decomposition to its first $r$ components assigns each Hankel row a reduced coordinate. In the reported experiments, these coordinates are the retained weighted-SVD row coordinates. With the reduced state denoted by $\x_i\in\R^r$, the controlled reduced-order model is
\begin{align}
    \x_{i+1} &= A\x_i+B\psivec_i,\label{eq:dmdc_state}\\
    \widehat{\y}_{i+q-1} &= C\x_i.\label{eq:dmdc_output}
\end{align}
The matrices $A$, $B$, and $C$ are identified from the reduced coordinates obtained on the training data. Arranging the $M$ training snapshots into consecutive one-step pairs, let
\begin{align}
    X_+ &= [\x_2,\x_3,\ldots,\x_M],\quad
    X_- = [\x_1,\x_2,\ldots,\x_{M-1}],\label{eq:Xpm}\\
    \Psi_- &= [\psivec_1,\psivec_2,\ldots,\psivec_{M-1}],\qquad
    \Theta = \begin{bmatrix}X_-\\\Psi_-\end{bmatrix}.\label{eq:theta}
\end{align}
Each column of $\Theta$ contains a reduced state and its corresponding command-feature vector, while the matching column of $X_+$ provides the reduced state at the next time step used to identify $A$ and $B$.

The pair $[A\;B]$ minimizes the weighted one-step error $\sum_{i} w_i\,\lVert\x_{i+1}-A\x_i-B\psivec_i\rVert_2^2$ under a ridge penalty $\lambda\lVert[A\;B]\rVert_F^2$, which has the closed form
\begin{equation}
    [A\;B]=X_+W_\Theta\Theta^{\top}
    \left(\Theta W_\Theta\Theta^{\top}+\lambda I\right)^{-1}.
    \label{eq:ridge}
\end{equation}
The diagonal matrix $W_\Theta=\diag(w_1,\ldots,w_{M-1})$ contains the weights from \eqref{eq:weights_general} restricted to the one-step regression pairs. Thus, the same event-sensitive weighting used to construct the reduced basis is also used to identify the reduced dynamics. The ridge term $\lambda I$ improves the conditioning of the weighted normal equations. The regularization parameter $\lambda$ is selected jointly with the delay length $q$ and retained rank $r$ using the transient-aware tuning criterion described next. The output map $C$ is obtained from a weighted least-squares fit from $\x_i$ to $\y_{i+q-1}$ under the same weights, allowing the reduced state to decode to terminal voltage and frequency.

After $[A\;B]$ is identified, a stability projection is applied to $A$. Eigenvalues with modulus $|\mu|>1$ are projected to a fixed radius $\rho<1$ while preserving their complex arguments. Complex-conjugate pairs are adjusted together so that the projected matrix remains real. The radius $\rho$ is specified a priori and is not tuned on the test interval. This projection ensures bounded recursive rollouts.

For online prediction, the latest measured delay window must be embedded into the same reduced coordinate system used during training. The row weighting $W_b^{1/2}$ in \eqref{eq:weighted_svd} is not applied to a live window since part of its construction depends on output-transient information unavailable in real time. Instead, the online encoder is the fixed projection onto the retained right-singular subspace,
\begin{equation}
	\x_k=E\,\hvec_k^{\leftarrow},\qquad E=V_r^{\top},
	\label{eq:encoder}
\end{equation}
where $V_r$ contains the first $r$ right singular vectors from \eqref{eq:weighted_svd} and $\hvec_k^{\leftarrow}$ uses the same per-channel standardization as the training Hankel windows. The same encoder is used for tuning, calibration, and testing. During rollout, $\x_{i+\ell}$ is propagated by \eqref{eq:dmdc_state} under the planned command features, and $C\x_{i+\ell}$ gives the predicted output at time $i+q-1+\ell$.

\subsection{Transient-Aware Model Selection}
The final model configuration is selected on a separate tuning interval using a loss that balances accuracy across the tuning data with normalized transient-focused error measures:
\begin{align}
	\mathcal L_{\mathrm{tune}} ={}& \beta_g\left(\mathrm{RMSE}_V+\mathrm{RMSE}_f\right)\nonumber\\
	&+\beta_e\left(\mathrm{RMSE}_{V,e}+\mathrm{RMSE}_{f,e}\right)\nonumber\\
	&+\beta_p\left(\mathrm{RMSE}_{V,p}+\mathrm{RMSE}_{f,p}\right)\nonumber\\
	&+\beta_\Delta\left(\mathrm{RMSE}_{\Delta V}+\mathrm{RMSE}_{\Delta f}\right).
	\label{eq:validation_loss}
\end{align}
Here, $\mathrm{RMSE}_V$ and $\mathrm{RMSE}_f$ are the full-interval voltage and frequency errors, $\mathrm{RMSE}_{V,e}$ and $\mathrm{RMSE}_{f,e}$ are event-weighted errors, $\mathrm{RMSE}_{V,p}$ and $\mathrm{RMSE}_{f,p}$ are peak-weighted errors, and $\mathrm{RMSE}_{\Delta V}$ and $\mathrm{RMSE}_{\Delta f}$ measure derivative-tracking error. The coefficients $\beta_g$, $\beta_e$, $\beta_p$, and $\beta_\Delta$ set the relative weights of global accuracy, command-transition accuracy, large-excursion accuracy, and response-shape accuracy. This criterion favors models that remain accurate during transient response intervals, not only during steady operation. Algorithm~\ref{alg:main} summarizes the deterministic forecasting procedure, including preprocessing, command-feature construction, event-weighted basis identification, reduced-order regression, model selection, and online rollout. 

\begin{algorithm}[b]
\caption{Event-Conditioned Hankel-DMDc:\\ \textit{offline identification and online open-loop forecast}}
\label{alg:main}
\small
\begin{algorithmic}[1]
\Require Training and tuning measurements; commanded $P/Q$ profiles; rollout horizon $H$
\Ensure Identified $(A,B,C,E)$; online open-loop forecast $\widehat{\y}_\ell$
\Statex \textbf{Offline: identification and selection}
\State Preprocess $V/f$ and scale commanded $P/Q$ with training-set statistics.
\State Build command-event features $\phivec_k$, event-memory channels, and the time-lagged command window $\psivec_i$.
\State Form training Hankel windows; compute basis weights $w_i$.
\State Compute the weighted SVD; retain rank $r$ and set $E=V_r^{\top}$.
\State Fit $A,B$ by weighted ridge and $C$ by weighted least squares; project $A$ for stability.
\State Select $(q,r,\lambda)$ on the tuning interval via $\mathcal{L}_{\mathrm{tune}}$.
\Statex \textbf{Online: open-loop look-ahead forecast}
\State Initialize $\x_0=E\hvec^{\leftarrow}$ from the pre-event window.
\For{$\ell=1,\dots,H$}
  \State $\x_\ell=A\x_{\ell-1}+B\psivec_{\ell-1}$,\quad $\widehat{\y}_\ell=C\x_\ell$
\EndFor
\end{algorithmic}
\end{algorithm}

%% file: sections/04_hardware_protocol.tex
\vspace{-2mm}
\section{Hardware Data and Evaluation Protocol}
\label{sec:data}
To evaluate the surrogate model, hardware data were collected from a testbed centered on a low-inertia synchronous generator. After preprocessing and normalization, the measured terminal signals and commanded active/reactive power profiles define the voltage, frequency, and command coordinates used for identification, calibration, and evaluation.
\vspace{-2mm}
\subsection{Hardware Setup and Measured Generator Profile}
\begin{figure}[t]
	\centering
	\includegraphics[width=0.68\linewidth]{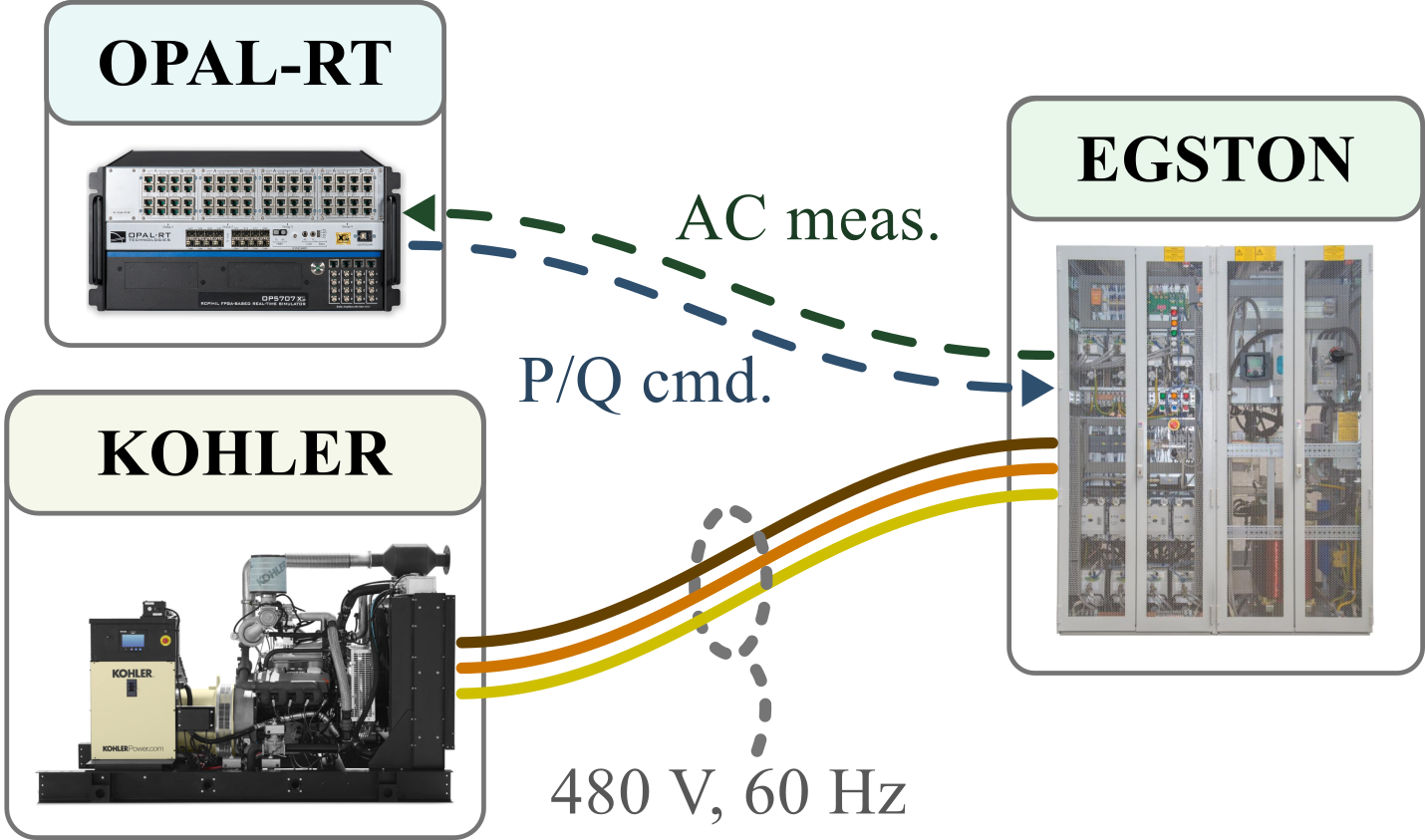}
	\caption{Experimental synchronous-generator testbed used for data collection. The EGSTON power amplifier serves as a programmable $P/Q$ load commanded by the OPAL-RT, which also records terminal measurements.}
	\label{fig:testbed}
\end{figure}

Fig.~\ref{fig:testbed} shows the synchronous-generator testbed used to collect the experimental data. The setup is centered around a Kohler KG180 generator rated at 180~kW/225~kVA, 480~V line-to-line, and 60~Hz. An EGSTON four-quadrant power amplifier provides programmable loading, while an OPAL-RT real-time simulator supplies the active- and reactive-power commands and records the resulting experimental data.

The command profile spans approximately one hour and consists of randomly generated active-power commands from 0 to 49~kW and reactive-power commands from 0 to 18~kVAr. The maximum apparent power was limited to 50~kVA by the available electrical infrastructure, while the power factor was varied from 0.8 lagging to unity. Terminal voltages and currents are measured internally by the amplifier and frequency is estimated from the voltages using a synchronous-reference-frame phase-locked loop (PLL). Because the command profile is known by construction, the commanded active and reactive powers are used as the exogenous inputs for forecasting.

\vspace{-2mm}
\subsection{Normalized Coordinates and Preprocessing}
The preprocessing stage converts the logged measurements into the RMS voltage, frequency, and command signals used for identification. Raw electrical channels are first processed at 2000~Hz and downsampled by a factor of 10 after anti-alias filtering, giving a 200~Hz model update rate and a discrete-time step of $\Delta t=0.005$~s. Voltage is represented by a three-cycle RMS estimate, while frequency is obtained using a PLL with a 5-Hz bandwidth. Hampel cleanup and target low-pass filters, with 12-Hz and 6-Hz cutoffs for voltage and frequency, respectively, are then applied before model identification.

These filtering steps are intended to suppress estimator ripple and high-frequency measurement artifacts, not to remove generator dynamics. Steady-state load intervals in this profile contain low-frequency oscillatory voltage and frequency behavior, on the order of 3~V and 0.5~Hz, which is retained in the measured hardware response. These oscillations originate in the low-inertia, engine-speed-governed dynamics of the generator rather than in the measurement chain.

The processed signals are then expressed in normalized coordinates for identification and calibration. At sample $k$, let $V_{\mathrm{rms},k}$ and $f_k$ denote the measured aggregate phase RMS voltage and frequency, and let $P_k$ and $Q_k$ denote the commanded active and reactive power. Using $V_{\mathrm{nom}}=277$~V, $f_{\mathrm{nom}}=60$~Hz, and $S_{\mathrm{base}}=225$~kVA, the normalized output and input coordinates are
\begin{equation} \y_k=\begin{bmatrix} (V_{\mathrm{rms},k}-V_{\mathrm{nom}})/V_{\mathrm{nom}}\\[1mm] (f_k-f_{\mathrm{nom}})/f_{\mathrm{nom}} \end{bmatrix}, \quad \uvec_k=\begin{bmatrix}P_k/S_{\mathrm{base}}\\Q_k/S_{\mathrm{base}}\end{bmatrix}. \label{eq:normalized_coords} \end{equation}

Although identification is performed in normalized coordinates, prediction errors are reported in engineering units. Let $y_{V,k}$ and $y_{f,k}$ denote the voltage and frequency components of $\y_k$, and let $\widehat{y}_{V,k}$ and $\widehat{y}_{f,k}$ denote the corresponding predicted components. The physical prediction errors are
\begin{align}
	\varepsilon_k^V &= V_{\mathrm{nom}}\bigl|\widehat y_{V,k}-y_{V,k}\bigr|,\label{eq:v_error_phys}\\
	\varepsilon_k^f &= f_{\mathrm{nom}}\bigl|\widehat y_{f,k}-y_{f,k}\bigr|.\label{eq:f_error_phys}
\end{align}
This convention keeps the regression well scaled while making the reported errors directly interpretable as volts and hertz.

All preprocessing stages are causal, making the framework realizable online without future samples. Together, the stages add about 100~ms of group delay, dominated by the three-cycle RMS voltage estimate, approximately 25~ms, and the frequency and target low-pass filters.
\vspace{-2mm}
\subsection{Data and Metrics}
The identification, tuning, and calibration data are taken from one continuous hardware profile over the intervals $[0,2400)$, $[2400,2700)$, and $[2700,3000)$~s, respectively. At the 200~Hz model rate, these intervals contain approximately 480{,}000, 60{,}000, and 60{,}000 samples and 381, 52, and 48 detected command events. Held-out testing uses a separate experimental run over $[3000,3600]$~s containing approximately 120{,}000 samples and 96 events. The test run does not influence model identification, tuning, or calibration.

Events are detected when the normalized command-change magnitude $m_k$ exceeds 5\% of its 95th-percentile training-set scale. Each event is labeled according to the signs of the active- and reactive-power changes, producing four joint step/shed event types, with a fifth label representing steady operation. The same labels are used for conformal direction stratification.

Deterministic prediction accuracy is evaluated using global, event-weighted, peak-weighted, and derivative RMSE for both voltage and frequency. Global RMSE measures overall tracking accuracy, while event-weighted RMSE emphasizes samples following commanded changes and peak-weighted RMSE emphasizes the largest observed deviations. Derivative RMSE measures agreement in the voltage and frequency rates of change. For each detected event and forecast horizon, nadir error is also computed as the absolute difference between the measured and predicted minimum voltage or frequency.

Per-event look-ahead forecasts are initialized from the measured pre-event window and propagated open-loop under the known command trajectory, with no measured voltage or frequency fed back during the forecast.

%% file: sections/05_results_deterministic.tex
\vspace{-2mm}
\section{Deterministic Prediction Results}
\label{sec:results_det}
The deterministic prediction results are presented next. The analysis begins with model-order selection, then evaluates accuracy, look-ahead behavior, trajectory tracking, and model variants used to assess the contribution of event conditioning.

\subsection{Hankel Diagnostic and Model Selection}
\begin{figure}[t]
	\vspace{0mm}
	\centering
	\includegraphics[width=0.65\linewidth]{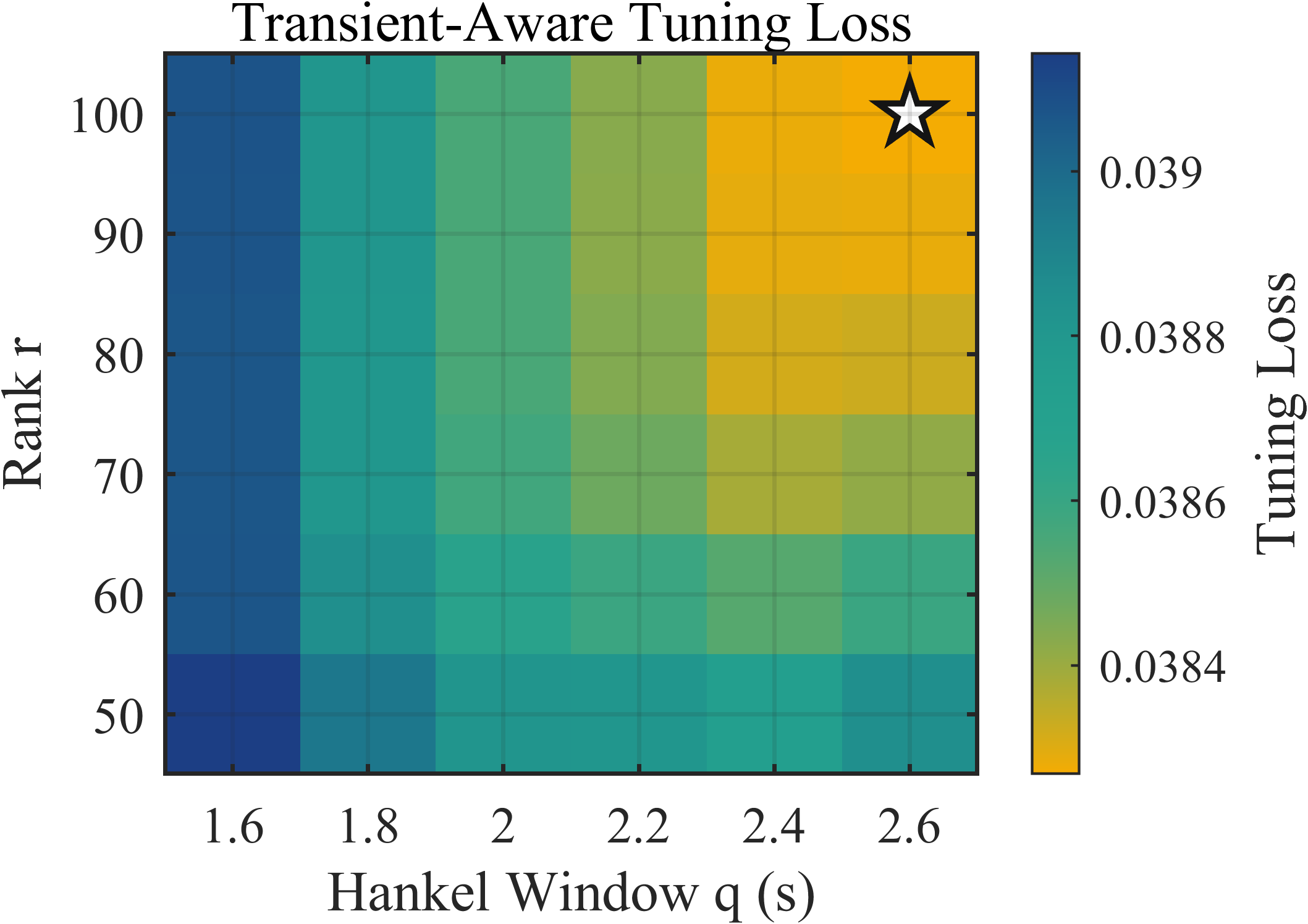}
	\caption{Transient-aware tuning loss over the Hankel-window and rank grid. The selected configuration, $q=520$, $r=100$, and $\lambda=1$, is marked.}
	\label{fig:validation}
\end{figure}

The final hyperparameters were selected on the tuning interval by searching over the Hankel-window length $q$, rank $r$, and ridge parameter $\lambda$. The transient-aware tuning loss selected $q=520$ samples (2.6 s), $r=100$, and $\lambda=1$. As shown in Fig.~\ref{fig:validation}, the loss decreases toward longer windows and higher ranks, placing the selected model at the longest-window, highest-rank corner of the tested grid. Because the surface is shallow, varying by less than 3\% across the grid, this selection should be interpreted as the best within the tested range rather than as a sharply bracketed interior optimum.

The remaining design choices were fixed in advance. These include command-feature stride $s=4$ ($L=131$ lags, giving a 1{,}572-dimensional command-feature vector), event-memory time constants $\tau\in\{0.05,0.15,0.50,1.50\}$ s, basis-weight coefficients $\alpha_u=8$, $\alpha_y=5$, and $\alpha_s=3$ with cap $w_{\max}=25$, tuning-loss weights $\beta_g=0.5$, $\beta_e=4$, $\beta_p=2$, and $\beta_\Delta=1.5$, and stability radius $\rho=0.98$.

\begin{figure}[b]
	\centering
	\includegraphics[width=0.82\linewidth]{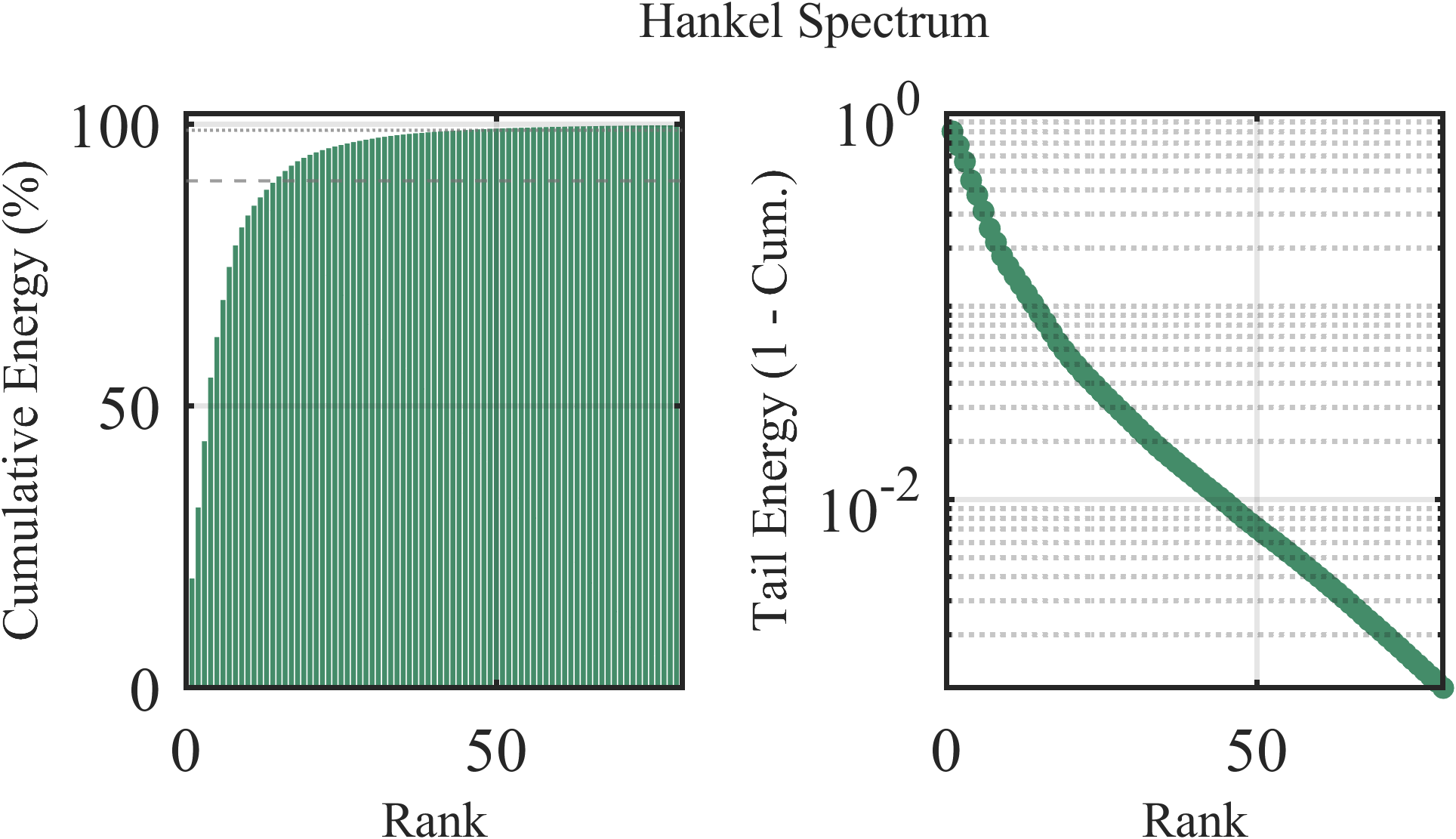}
	\caption{Hankel singular-value diagnostic at the selected window length $q=520$. Left: cumulative energy versus rank. Right: tail energy versus rank.}
	\label{fig:hankel_diag}
\end{figure}

Fig.~\ref{fig:hankel_diag} shows the Hankel singular-value diagnostic at the selected window length $q=520$. For this window, the training-interval Hankel matrix has 479{,}481 rows and 1{,}040 columns. The singular-value energy is highly concentrated, with ranks 15 and 45 capturing 90\% and 99\% of the energy, respectively. The effective rank is 15.85 and the stable rank is 5.14, which indicates compact delay-coordinate structure. However, the non-negligible tail suggests that sparse transient behavior is not fully captured by the leading global-energy directions alone. Retaining $r=100$ modes therefore provides additional capacity for transient response components while preserving a reduced-order representation.

\vspace{-2mm}
\subsection{Deterministic Prediction Accuracy}
\begin{figure*}[!b]
	\centering
	\includegraphics[width=0.8\textwidth]{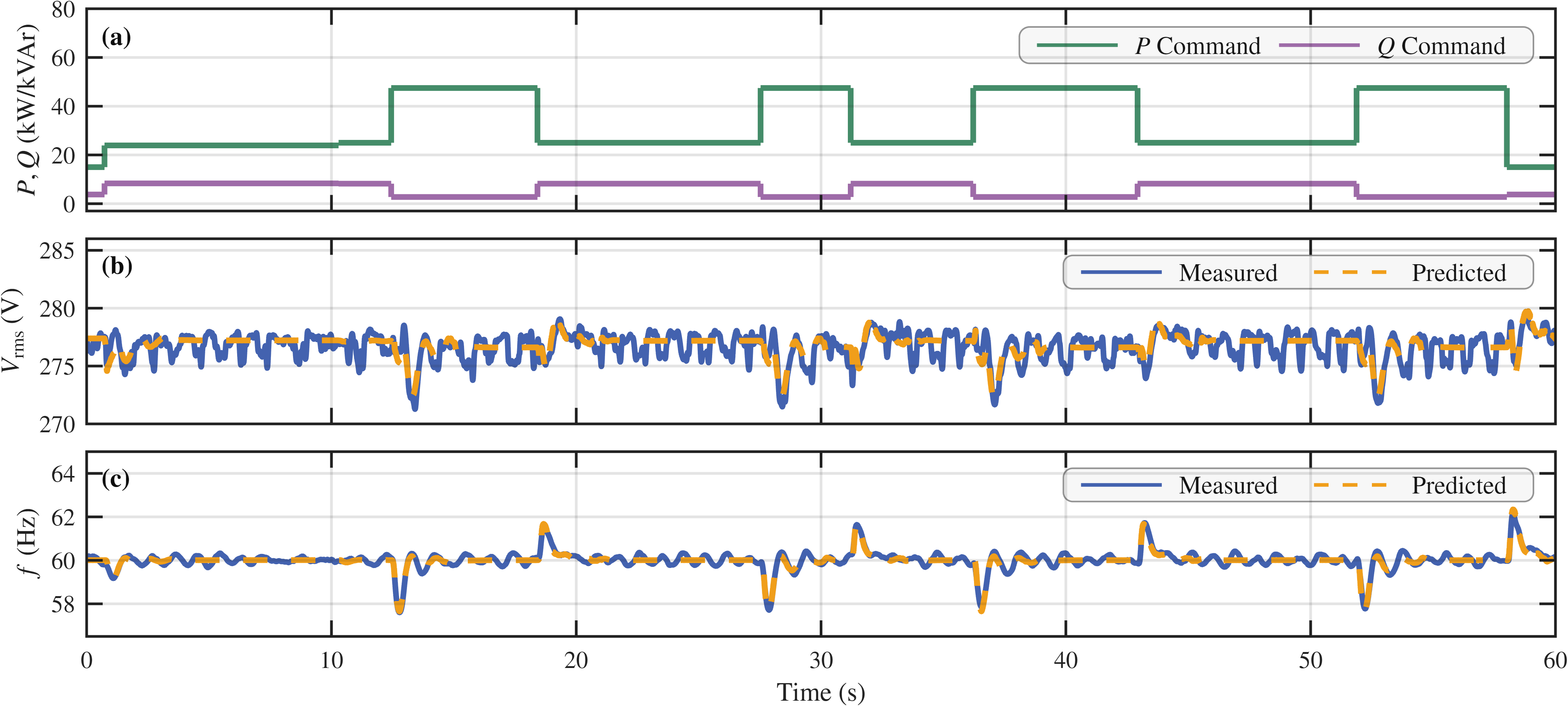}
	\caption{Representative one-minute held-out interval. Panel (a) shows the commanded active and reactive load, panel (b) shows the measured and predicted RMS voltage response, and panel (c) shows the measured and predicted frequency response.}
	\label{fig:deterministic_zoom}
\end{figure*}

Table~\ref{tab:deterministic} reports the accuracy of the model. Long steady intervals dominate the load profile, so the global RMSE largely reflects steady tracking and provides an overall accuracy check. The event-weighted and per-event errors provide the main transient-focused measures, while the peak-weighted, derivative, and per-event nadir metrics in Tables~\ref{tab:deterministic} and~\ref{tab:lookahead_acc} are more relevant for screening against voltage and frequency limits since they reflect response extremes and recovery shape. On the test interval, the event-weighted RMSE is 1.058~V and 0.155~Hz, compared with global RMSE values of 0.954~V and 0.150~Hz. The contribution of event conditioning is assessed separately through the model-variant comparison in Table~\ref{tab:ablation_plan}.
\begin{table}[t]
\caption{Deterministic Prediction Accuracy.}
\label{tab:deterministic}
\centering
\scriptsize
\setlength{\tabcolsep}{4pt}
\begin{tabular}{@{}lrrr@{}}
\toprule
Metric (RMSE) & Train & Validation & Test \\
\midrule
Global voltage (V) & 0.868 & 0.900 & 0.954 \\
Global frequency (Hz) & 0.143 & 0.142 & 0.150 \\
Event-window voltage (V) & 0.911 & 0.938 & 1.058 \\
Event-window frequency (Hz) & 0.145 & 0.144 & 0.155 \\
Peak-weighted voltage (V) & 0.935 & 0.966 & 1.061 \\
Peak-weighted frequency (Hz) & 0.145 & 0.147 & 0.157 \\
Voltage-rate  $dV/dt$ (V/s) & 13.03 & 13.18 & 16.44 \\
Frequency-rate  $df/dt$ (Hz/s) & 1.77 & 1.72 & 1.62 \\
\bottomrule
\end{tabular}
\end{table}
\begin{table}[t]
	\caption{Per-Event Look-Ahead Accuracy by Horizon.}
	\label{tab:lookahead_acc}
	\centering
	\scriptsize
	\setlength{\tabcolsep}{4pt}
	\begin{tabular}{@{}lrrrr@{}}
		\toprule
		Horizon & Event V & Event f & V nadir & f nadir \\
		& RMSE (V) & RMSE (Hz) & err. (V) & err. (Hz) \\
		\midrule
		0.5 s & 1.511 & 0.192 & 0.859 & 0.128 \\
		1.0 s & 1.303 & 0.196 & 1.243 & 0.126 \\
		2.0 s & 1.137 & 0.175 & 1.540 & 0.139 \\
		5.0 s & 1.007 & 0.159 & 1.649 & 0.181 \\
		\bottomrule
	\end{tabular}
\end{table}
Accuracy remains consistent across training, the combined validation and calibration interval, and the held-out test. This consistency is important because the held-out test set comes from a separate experimental run. Fig.~\ref{fig:deterministic_zoom} shows a representative one-minute segment from the held-out test interval.
\vspace{-2mm}
\subsection{Look-Ahead Accuracy by Horizon}
Table~\ref{tab:lookahead_acc} reports per-event open-loop prediction accuracy on the held-out test data at fixed post-event horizons. The event-weighted RMSE remains low across the tested horizons and decreases slightly as longer windows include more settled response samples after the initial transient. The mean absolute nadir error increases with horizon, reflecting the greater likelihood that longer windows include the most severe voltage or frequency deviation. These results show that average trajectory accuracy and transient-response accuracy capture different aspects of look-ahead performance.


\vspace{-2mm}
\subsection{Model-Variant Comparison}
Table~\ref{tab:ablation_plan} compares model variants on the held-out test interval. Each variant is retrained using the final hyperparameters and evaluated using two metrics: global RMSE over the full profile and per-event open-loop look-ahead RMSE over a 2-s post-event window. The largest performance gain comes from the event conditioning provided by the command-event features. Relative to the same-preprocessing no-event model, event features reduce the per-event voltage RMSE from 1.077~V to 0.967~V. The event-weighted basis has a more targeted effect. It does not reduce the per-event window RMSE relative to the features-only model, with the full model giving 1.017~V compared with 0.967~V. Its strongest effect appears in the voltage nadir metric, where the basis-only variant gives the lowest error, 1.523~V, while the combined model remains close at 1.529~V. Therefore, the event-weighted basis is retained as a transient-fidelity mechanism, while the command-event features account for the primary improvement in overall accuracy. The stability projection has no effect on these results because the fitted dynamics are already stable, giving identical metrics with and without the projection.
\begin{table*}[t]
	\caption{Model-Variant (Ablation) Comparison on the Held-Out Test Interval.}
	\label{tab:ablation_plan}
	\centering
	\scriptsize
	\setlength{\tabcolsep}{5pt}
	\begin{tabular}{@{}lrrrrrr@{}}
		\toprule
		Variant & \shortstack{Global V\\(V)} & \shortstack{Global f\\(Hz)} & \shortstack{Per-event\\win. V (V)} & \shortstack{Per-event\\win. f (Hz)} & \shortstack{Per-event\\V nadir (V)} & \shortstack{Per-event\\f nadir (Hz)} \\
		\midrule
		Same-preprocessing Hankel-DMDc (no events) & 0.948 & 0.165 & 1.077 & 0.192 & 1.611 & 0.139 \\
		\;+ event features only & 0.922 & 0.145 & 0.967 & 0.148 & 1.537 & 0.136 \\
		\;+ event-weighted basis only & 0.955 & 0.166 & 1.088 & 0.194 & 1.523 & 0.134 \\
		\;+ event features and weighting (full) & 0.954 & 0.150 & 1.017 & 0.160 & 1.529 & 0.137 \\
		Full model, no stability projection & 0.954 & 0.150 & 1.017 & 0.160 & 1.529 & 0.137 \\
		\bottomrule
	\end{tabular}
\end{table*}

%% file: sections/06_conformal_reliability.tex
\section{Conformal Prediction Bands}
\label{sec:conformal}

Because the residuals of a data-driven generator surrogate model do not need to follow a prescribed parametric distribution, and because errors widen around command transitions, the frozen surrogate model is paired with an event-conditioned split-conformal calibration layer. This layer uses held-out residuals to construct joint voltage/frequency bands whose widths adapt to command-event magnitude, phase, and direction across the forecast horizon.
\vspace{-3mm}
\subsection{Calibration Principle}
Split conformal regression uses held-out residuals to construct prediction sets with finite-sample marginal coverage under exchangeability \cite{vovk2005,lei2018}. Related methods, including conformalized quantile regression and sequential conformal prediction, address heterogeneous or drifting residual distributions in broader settings \cite{romano2019,xu2023}. Here, calibration is conditioned on command-event attributes so that the band width is estimated from held-out errors observed under similar operating conditions. After the deterministic model is selected and frozen, calibration modifies only the prediction bands and not the point forecast. For each calibration sample $k\in\Ical$, the absolute voltage and frequency errors are
\begin{equation}
    e^V_k=|\widehat y_{V,k}-y_{V,k}|,\qquad e^f_k=|\widehat y_{f,k}-y_{f,k}|.
    \label{eq:residuals}
\end{equation}
Given a finite set of scores $\mathcal S=\{s_1,\ldots,s_n\}$, written in sorted order $s_{(1)}\leq\cdots\leq s_{(n)}$, the split-conformal quantile is
\begin{equation}
    Q_{1-\alpha}(\mathcal S)=s_{(k_\alpha)},\qquad
    k_\alpha=\min\{\lceil(n+1)(1-\alpha)\rceil,n\}.
    \label{eq:conf_quantile}
\end{equation}
The level $\alpha$ is the conformal miscoverage rate, distinct from the event-weight coefficients $\alpha_u,\alpha_y,\alpha_s$ in Section~\ref{sec:method}. Under exchangeability within a fixed context, and with both the surrogate model and context map fixed before calibration, \eqref{eq:conf_quantile} gives marginal coverage of at least $1-\alpha$ for the corresponding prediction set. In the present time-series setting, this guarantee is used as a calibration principle and is verified by held-out empirical coverage, since neighboring errors and errors within the same event are correlated.
\vspace{-3mm}
\subsection{Event-Conditioned Joint Bands}

Each sample is assigned an event context derived from the active- and reactive-power command trajectory:
\begin{equation}
    c_k=\left(b_k^{\gamma},b_k^{\theta},b_k^{\delta}\right),
    \label{eq:context}
\end{equation}
where $b_k^{\gamma}$ bins event magnitude, $b_k^{\theta}$ bins time since the most recent event, and $b_k^{\delta}$ labels the joint active/reactive event direction. The phase bins span the immediate post-event response through recovery, allowing the band to follow the transient error envelope. Direction labels include a steady label and four joint $P/Q$ event types corresponding to step or shed behavior in each command channel.

Fine contexts are used when sufficiently supported by calibration data, otherwise the radius backs off to magnitude-phase, magnitude-only, and finally the global calibration pool. Therefore, each band is sized from the most specific calibration pool supported by the data. A fixed inflation factor slightly greater than one is applied to every context radius to provide conservative margin for temporal dependence and smoothing effects, and radii are smoothed using only past samples through a one-sided causal window so that the bands can be produced online.

Operating decisions depend on voltage and frequency simultaneously, so the bands are calibrated jointly. For each context $c$, preliminary channel scales $r_V(c)$ and $r_f(c)$ are obtained from calibration-error quantiles using the same fallback hierarchy. The joint score then standardizes both errors and uses the larger standardized miss:
\begin{equation}
    s_k^J=\max\left\{\frac{e_k^V}{r_V(c_k)+\epsilon},\frac{e_k^f}{r_f(c_k)+\epsilon}\right\}.
    \label{eq:joint_score}
\end{equation}
The conformal quantile of the joint scores within a context gives the multiplier
\begin{equation}
    \rho_J(c)=Q_{1-\alpha}\left(\{s_k^J:k\in\Ical,\ c_k=c\}\right),
    \label{eq:joint_multiplier}
\end{equation}
with fallback to coarser pools when support is sparse. The resulting joint prediction set is
\begin{equation}
    \begin{split}
        \mathcal C_k^J=\{(y_V,y_f):&|y_V-\widehat y_{V,k}|\leq \rho_J(c_k)r_V(c_k),\\
        &|y_f-\widehat y_{f,k}|\leq \rho_J(c_k)r_f(c_k)\}.
    \end{split}
    \label{eq:joint_band}
\end{equation}
Although voltage and frequency bands are plotted separately, \eqref{eq:joint_band} is calibrated as one joint object. A sample is counted as covered only when both measured quantities lie inside their corresponding intervals at the same instant. For any region $\mathcal R$, the empirical pointwise joint coverage is
\begin{equation}
    \widehat{\mathrm{Cov}}^{J}(\mathcal R)=\frac{1}{|\mathcal R|}\sum_{k\in\mathcal R}
    \mathbb I\{y_{V,k}\in\mathcal C_{k,V}^{J}\}\mathbb I\{y_{f,k}\in\mathcal C_{k,f}^{J}\}.
    \label{eq:coverage}
\end{equation}
An event-weighted version replaces the uniform average with command-event weights. The coverage in \eqref{eq:coverage} is pointwise. It tests simultaneous voltage/frequency inclusion sample by sample, not containment of an entire event trajectory.
\vspace{-2mm}
\subsection{Held-Out Coverage Results}
With the calibration layer fixed, the bands are evaluated on the same held-out test interval used for the deterministic results. All bands target 90\% nominal coverage, so $\alpha=0.10$. The event-context grid combines four event-magnitude bins, seven event-phase bins split at 0.05, 0.15, 0.30, 0.70, 1.50, and 3.00~s, and five direction labels consisting of the steady label and the four joint $P/Q$ event types. A fine context is used only when it contains at least 120 calibration samples, with the event-only fallback requiring 250 samples. Otherwise, calibration backs off through the hierarchy described above. The resulting context radii are inflated by 1.05 and smoothed with a 0.10-s causal window.

\begin{figure*}[!t]
    \centering
    \includegraphics[width=0.82\textwidth]{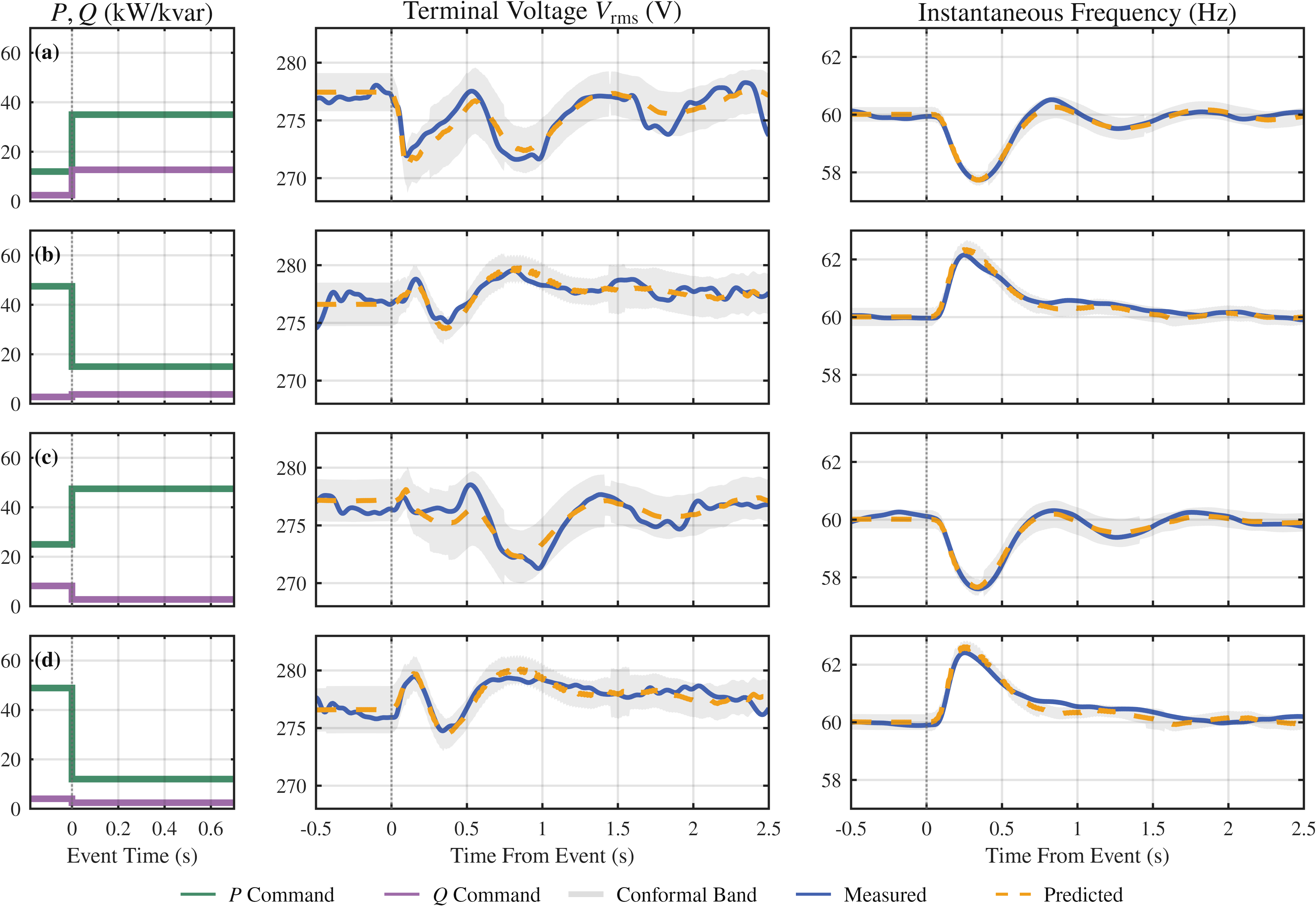}
    \caption{Representative held-out test events with conformal prediction bands. Rows show four joint command-event types: (a) $P$ step, $Q$ step; (b) $P$ shed, $Q$ step; (c) $P$ step, $Q$ shed; and (d) $P$ shed, $Q$ shed. In each row, the left panel shows the commanded $P/Q$ change, the middle panel shows measured and predicted voltage, and the right panel shows measured and predicted frequency.}
    \label{fig:event_zooms}
\end{figure*}

Fig.~\ref{fig:event_zooms} shows representative held-out events with joint conformal bands. In each case, the point forecast tracks the local voltage/frequency response while the band is intended to contain both measured channels through the transient and recovery. Table~\ref{tab:coverage} reports empirical joint coverage and mean band width by region. On the held-out test data, joint coverage is 90.17\%, closely matching the 90\% target. Event-weighted joint coverage is similarly close at 90.08\%, indicating reliable coverage during command-transition samples as well as steady operation. The mean held-out band widths are 3.55~V and 0.566~Hz. The training row is included only as a diagnostic, while the calibration and held-out test rows carry the main empirical coverage results.


\begin{table}[b]
    \caption{Empirical Joint Coverage and Mean Band Width.}
    \label{tab:coverage}
    \centering
    \scriptsize
    \setlength{\tabcolsep}{3.5pt}
    \begin{tabular}{@{}lcrrrr@{}}
	        \toprule
	        Region & $n$ & Joint cov. & \shortstack{Event-joint} & \shortstack{Mean V width} & \shortstack{Mean f width} \\
	        & & (\%) & (\%) & (V) & (Hz) \\
	        \midrule
	        Train  & 479{,}481 & 91.31 & 91.11 & 3.58 & 0.556 \\
	        Validation & 119{,}481 & 91.96 & 92.07 & 3.61 & 0.557 \\
	        Calibration & 59{,}481 & 92.86 & 93.48 & 3.61 & 0.559 \\
	        Held-out test & 119{,}481 & 90.17 & 90.08 & 3.55 & 0.566 \\
	        \bottomrule
	    \end{tabular}
\end{table}

Held-out coverage resolved by event phase was at or above target immediately after command events, where the bands are widest, and slightly lower during later recovery as the bands narrow. Table~\ref{tab:event_type_coverage} reports held-out coverage for the four joint command-event types. The lowest-coverage case is the ($P$ shed, $Q$ step) event type, which reaches 83.12\% joint coverage and has both the fewest test events and the narrowest bands. This shortfall is therefore most consistent with limited calibration support in that region of the event space.

%% file: sections/09_discussion.tex
\section{Discussion and Deployment Scope}
\label{sec:discussion}

The experiments demonstrate an event-conditioned surrogate model for generator transient forecasting with calibrated joint voltage/frequency prediction bands. After training, the model uses a 100-dimensional reduced state, a 1{,}572-entry command-feature window, and 2.6~s of measured history. On a single core of an Intel Core Ultra 7 155H, a 1-s forecast requires 5.28~ms and a 50-s forecast 181~ms, approximately 280 times faster than real time, with about 15~MB of memory. The primary computational cost is offline basis construction at 1{,}057~s, compared with 10.1~s for preprocessing, 359~s for fitting and tuning, and 2.9~s for calibration.

The evaluation is limited to data from the same physical generator, although the held-out test set comes from a separate experimental run. The results therefore demonstrate transfer to a new operating profile but not generalization to different generators, controllers, sensor chains, or load magnitudes.

The open-loop rollout can accumulate error with horizon, as reflected by the nadir metrics in Table~\ref{tab:lookahead_acc}, so operational use should consider the calibrated band rather than the point forecast alone. The three-cycle RMS estimate and 5-Hz frequency tracker limit the modeled response to the sub-ten-hertz envelope rather than fast electromagnetic transients. The method assumes the future command is known, so the bands quantify residual model error under calibrated command and state contexts rather than command uncertainty.

\begin{table}[b]
	\caption{Joint Conformal Coverage by Event Type (Held-Out Test).}
	\label{tab:event_type_coverage}
	\centering
	\scriptsize
	\setlength{\tabcolsep}{2.4pt}
	\begin{tabular}{@{}lrrrrr@{}}
		\toprule
		Event type & Events & Joint cov. & Event-joint & Mean V width & Mean f width \\
		& & (\%) & (\%) & (V) & (Hz) \\
		\midrule
		$P$ step, $Q$ step & 31 & 90.66 & 89.30 & 4.05 & 0.589 \\
		$P$ step, $Q$ shed & 22 & 92.76 & 93.47 & 3.96 & 0.682 \\
		$P$ shed, $Q$ step & 15 & 83.12 & 84.81 & 3.05 & 0.519 \\
		$P$ shed, $Q$ shed & 28 & 92.22 & 91.64 & 3.10 & 0.496 \\
		\bottomrule
	\end{tabular}
\end{table}

The measured hardware exhibits steady-state voltage and frequency oscillations of approximately 3~V and 0.5~Hz, which limit the minimum achievable band width. These oscillations are associated with the low-inertia gaseous-engine generator dynamics rather than measurement noise.

The surrogate model is intended as an advisory screening tool rather than a replacement for protection settings or validated stability studies. Commands outside the calibrated magnitude, phase, or direction support should be treated as out of support. The bands should be recalibrated as operating conditions change, and the surrogate model should be updated when the asset, controller, or measurement chain changes. Operational screening should compare the joint prediction band against system-specific voltage and frequency limits. If the band crosses a limit, the command should be flagged for caution or higher-fidelity validation.

%% file: sections/11_conclusion.tex
\section{Conclusion}
\label{sec:conclusion}

In this paper, an event-conditioned Hankel-DMDc surrogate model was developed for generator-DT transient forecasting under planned active- and reactive-power load commands. The model combines delay-coordinate lifting, command-event features, and an event-weighted Hankel basis to predict terminal voltage and frequency. Hardware experiments on a low-inertia synchronous-generator testbed showed accurate transient tracking on a held-out experimental run while maintaining computation times suitable for online screening.

A conformal calibration layer augments the frozen surrogate model with event-conditioned pointwise joint voltage/frequency prediction bands. On the held-out test interval, the bands achieved near-nominal joint coverage and provide uncertainty bounds for operating-limit screening. The framework is intended for calibrated command and state contexts, with limitations related to calibration support, open-loop forecast horizon, measurement bandwidth, and testbed operating range. Future work will extend validation to additional generators and operating conditions and investigate trajectory-level conformal guarantees.\vspace{-2.5mm}

%% file: sections/12_acknowledgment.tex
\section*{Acknowledgment}
This work was supported by the Office of Naval Research under contracts No.~N00014-23-C-1012 and No.~N00014-24-C-1301. DISTRIBUTION STATEMENT A. Approved for public release: distribution is unlimited. Approved, DCN\# 2026-6-30-2452.